\documentclass{emulateapj}

\shorttitle{Fermi observations of GRB~080825C}
\shortauthors{Fermi GBM \& LAT Teams}

\begin{document}

\title{Fermi observations of high-energy gamma-ray emission from GRB~080825C}

\author{
A.~A.~Abdo\altaffilmark{2,3}, 
M.~Ackermann\altaffilmark{4}, 
K.~Asano\altaffilmark{5,6}, 
W.~B.~Atwood\altaffilmark{7}, 
M.~Axelsson\altaffilmark{8,9}, 
L.~Baldini\altaffilmark{10}, 
J.~Ballet\altaffilmark{11}, 
D.~L.~Band\altaffilmark{12,13,14}, 
G.~Barbiellini\altaffilmark{15,16}, 
D.~Bastieri\altaffilmark{17,18}
% ...
}
% ...
\altaffiltext{2}{Space Science Division, Naval Research Laboratory, Washington, DC 20375}
\altaffiltext{3}{National Research Council Research Associate, National Academy of Sciences, Washington, DC 20001}
\altaffiltext{4}{W. W. Hansen Experimental Physics Laboratory, Kavli Institute for Particle Astrophysics and Cosmology, Department of Physics and SLAC National Accelerator Laboratory, Stanford University, Stanford, CA 94305}
\altaffiltext{5}{Department of Physics, Tokyo Institute of Technology, Meguro City, Tokyo 152-8551, Japan}
\altaffiltext{6}{Interactive Research Center of Science, Tokyo Institute of Technology, Meguro City, Tokyo 152-8551, Japan}
\altaffiltext{7}{Santa Cruz Institute for Particle Physics, Department of Physics and Department of Astronomy and Astrophysics, University of California at Santa Cruz, Santa Cruz, CA 95064}
\altaffiltext{8}{Department of Astronomy, Stockholm University, SE-106 91 Stockholm, Sweden}
\altaffiltext{9}{The Oskar Klein Centre for Cosmo Particle Physics, AlbaNova, SE-106 91 Stockholm, Sweden}
\altaffiltext{10}{Istituto Nazionale di Fisica Nucleare, Sezione di Pisa, I-56127 Pisa, Italy}
\altaffiltext{11}{Laboratoire AIM, CEA-IRFU/CNRS/Universit\'e Paris Diderot, Service d'Astrophysique, CEA Saclay, 91191 Gif sur Yvette, France}
\altaffiltext{12}{NASA Goddard Space Flight Center, Greenbelt, MD 20771}
\altaffiltext{13}{Center for Research and Exploration in Space Science and Technology (CRESST), NASA Goddard Space Flight Center, Greenbelt, MD 20771}
\altaffiltext{14}{Deceased}
\altaffiltext{15}{Istituto Nazionale di Fisica Nucleare, Sezione di Trieste, I-34127 Trieste, Italy}
\altaffiltext{16}{Dipartimento di Fisica, Universit\`a di Trieste, I-34127 Trieste, Italy}
\altaffiltext{17}{Istituto Nazionale di Fisica Nucleare, Sezione di Padova, I-35131 Padova, Italy}
\altaffiltext{18}{Dipartimento di Fisica ``G. Galilei", Universit\`a di Padova, I-35131 Padova, Italy}
% ...

\begin{abstract}
The Fermi Gamma-ray Space Telescope (FGST) has opened a new
high-energy window in the study of Gamma-Ray Bursts (GRBs).  Here we
present a thorough analysis of GRB~080825C, which triggered the Fermi
Gamma-ray Burst Monitor (GBM), and was the first firm detection of a
GRB by the Fermi Large Area Telescope (LAT).  We discuss the LAT event
selections, background estimation, significance calculations, and
localization for Fermi GRBs in general and GRB~080825C in particular.
We show the results of temporal and time-resolved spectral analysis of
the GBM and LAT data. We also present some theoretical interpretation
of GRB~080825C observations as well as some common features observed
in other LAT GRBs.
\end{abstract}


\section{Introduction}

Gamma-Ray Bursts (GRBs) originate from the most luminous explosions in
the universe and more than 35 years after their
discovery in 1967 \citep{kel73}, many questions remain to be answered
about their possible progenitors, the composition of the
ultra-relativistic outflows that power them, and the dominant emission
mechanism for their prompt gamma rays.  The Burst And Transient Source
Experiment (BATSE) onboard the Compton Gamma-Ray Observatory (CGRO;
1991-2000) made significant advances to the field, thoroughly exploring
the 25~keV -- 2~MeV energy range with detailed population studies of the
prompt gamma-ray emission.  Burst spectra were found to be well
described by the Band function
\citep{band93}, which consists of two smoothly connected power laws.
It was understood, however, that observations of GRBs at higher energies were of crucial importance to
resolve some of the open issues: constrain the bulk
Lorentz factor of the outflow and the distance from the central source
to the gamma-ray emission region, distinguish between hadronic
and leptonic origins of the gamma-ray emission, and probe for
signatures of Ultra High Energy Cosmic Rays (UHECRs) which could be accelerated 
within GRB jets \citep[see][for a review of the prospects for GRB science with Fermi LAT]{band08}.

Constraints on the origin of the high-energy emission from GRBs are
quite limited due to both the small number of bursts with firm
high-energy detection and the small number of events that were
detected in such cases.  High-energy emission from GRBs was first
observed by the Energetic Gamma-Ray Experiment Telescope (EGRET,
covering the energy range from $30\;$MeV to $30\;$GeV) onboard
CGRO. Emission above $100\;$MeV was detected in five cases: 
GRBs~910503, 910601, 930131, 940217 and 940301 \citep{dingus95}. 
One of these sources, GRB~930131, had high-energy emission that was consistent with an
extrapolation from its spectrum obtained with BATSE between 25~keV --
4~MeV~\citep{som94}.  In contrast, evidence for an additional
high-energy component up to $200\;$MeV with a different temporal
behavior to the low-energy component was discovered in GRB~941017
\citep[in EGRET's calorimeter TASC; ][]{gon03}. The high-energy emission for the latter GRB lasted more
than 200 seconds with a single spectral component being ruled out.
A unique aspect of the high-energy emission in GRB~940217
was its duration, which lasted up to $\sim$$90$
minutes after the BATSE GRB trigger, including an 18 GeV photon
at $\sim$$75$ minutes post-trigger
\citep{hur94}. More recently, the GRID instrument onboard 
Astro-rivelatore Gamma a Immagini LEggero (AGILE) detected 10
high-energy events with energies up to $300\;$MeV from GRB 080514B, in
coincidence with its lower energy emission, with a significance of
$3.0\;\sigma$ \citep{giu08}.

The Fermi Gamma-ray Space Telescope was launched on June 11 2008 
and provides an unprecedented energy coverage and
sensitivity for the study of high-energy emission in GRBs. It is
composed of two instruments: the Gamma-ray Burst Monitor 
\citep[GBM;][]{meg08} and the Large Area Telescope \citep[LAT;][]{atw09}. 
The GBM covers the entire unocculted sky with 12 sodium iodide (NaI)
detectors with different orientation placed around the spacecraft and
covering an energy range from 8 keV to 1 MeV, and two bismuth
germanate (BGO) scintillators placed on opposite sides of the
spacecraft with energy coverage from 200 keV to 40 MeV. The LAT is
a pair conversion telescope made up of $4 \times 4$ arrays of silicon
strip trackers and cesium iodide calorimeter modules covered by a
segmented anti-coincidence detector designed to efficiently reject
charged particles. The energy coverage of the LAT
instrument ranges from 20 MeV to more than 300~GeV with a
field-of-view (FoV) of $\sim$$2.4$ steradians. Note that the LAT effective area is still non-zero even as far out as 70 degrees off-axis which allows the detection of bursts with such high incident angles.  
As of June 1 2009, 9 GRBs have been
detected by the LAT at energies above 100~MeV: GRB~080825C
\citep{bou08}, GRB~080916C \citep{abd09, tajima08}, GRB~081024B \citep{omo08}, 
GRB~081215A \citep{mcenery08}, GRB~090217 \citep{ohn09}, GRB~090323 \citep{ohnovdh09}, 
GRB~090328 \citep{mcenery09, cutini09}, GRB~090510 \citep{ohnopela09, omo09}. 
In this paper, we report the observations and analyses of gamma-ray emission from
GRB~080825C, the first GRB detected by both the GBM and the LAT
instruments. Section \ref{sec:observations} will present the GBM and
LAT observations along with the various methods used for data analysis, 
section \ref{sec:dataana} provides the results of 
detailed time-resolved spectroscopy, and section \ref{sec:discussion} discusses the theoretical 
interpretation of our observations and compares the properties of this event 
to the ones observed in some other LAT GRBs.

\section{Burst detection and localization} \label{sec:observations}

\subsection{GBM observations}

At 14:13:48 UTC on August 25 2008 ($T_0$), GRB~080825C
triggered the GBM flight software~\citep[trigger
241366429,][]{vanderhorst08}. On-ground analysis of the GBM data
localized the burst at Right Ascension (RA, J2000) = 232.2$^\circ$,
Declination (Dec, J2000) = -4.9$^\circ$, with a statistical
uncertainty of 1.5$^\circ$ at the $1$-$\sigma$ confidence level.  The
GBM on-ground localization placed this GRB at $\sim$$60^\circ$ from
the LAT boresight at the time of the trigger, at the edge of the LAT
FoV, where the effective area is a factor of $\sim$$3$ less than on axis.

The top two panels of Figure~\ref{lightcurves} show the background
subtracted light curves (see section \ref{sec:specana}) of the two
brightest NaI detectors (9 \& 10) and of the two BGO detectors. The
GRB exhibits a multiple peak structure with the two brightest peaks
seen right after onset.  The $T_{90}$ and $T_{50}$ durations of the
event \citep[time during which 90\% and 50\% of the event flux was
collected, cf.][]{ck93} were estimated to be (8-1000~keV) $\sim$$27\;$s 
and $\sim$$13\;$s, respectively.  Emission in the NaI and BGO
detectors becomes extremely weak after $\sim$$T_0+25\;$s.
However, emission is detected in the NaI scintillators up to $\sim$$35\;$s after
the trigger time with a $3.6\;\sigma$ significance in bin (e).

\subsection{LAT observations}

The LAT events detected close to the GBM position around the trigger
time are shown in the bottom panel of Figure~\ref{lightcurves} (see details below). The
LAT data show a count rate increase that is spatially and temporally
correlated with the GBM emission. We have performed a detailed analysis
of the significance of this detection. Details on this computation as
well as of the LAT data selection, background estimation and localization are given below.

\subsubsection{Event selection}\label{sect:events}

Most of the events detected by the LAT instrument are cosmic rays that need to be distinguished from any source of $\gamma$-ray
signal.  The broad range of LAT observations and analyses, from GRBs
to extended diffuse radiation, leads to different optimizations of the
event selections which have different rates of residual backgrounds (misclassified cosmic rays). The
LAT background rejection analysis has been constructed to allow analysis
classes to be optimized for specific science topics \citep{atw09}. In the
case of GRB observations, the relatively small region of the sky as
well as the very short time window allow the background rejection cuts
to be relaxed relative to an analysis of a diffuse source covering a
large portion of the sky over longer periods of time.  Indeed, the so-called
`diffuse' event class is most suited for studying faint sources (like
diffuse gamma-ray emission) with minimum background contamination. In the case of GRB 080825C, it was used to search for possible high-energy afterglow emission up to 13~ks after the burst trigger
(see section \ref{sec:afterglow}). On shorter timescales, a significant improvement in signal-to-noise ratio can be obtained by increasing the effective area while keeping the background rate at a reasonable level. The so-called 'transient' event class was developed for this specific purpose and is used for burst detection and localization.

Using Monte-Carlo simulations as well as real data input, the event
selection used for spectral analysis has been optimized compared to
previous analyses \citep{abd09} that uses the 'transient' class for the prompt emission analysis. 
Indeed, we found that cuts can be
loosened even further for typical burst duration since background
contamination is less of an issue for short time windows. For this purpose, a more relaxed event class, so-called 'S3', has been developed to improve the LAT effective area at the expense of an increase in background rate. Pure effective area increase is on the order of $\sim$30 \% at 100 MeV and $\sim$10\% at 1 GeV.
For a typical GRB spectrum, the gamma efficiency above 80 MeV increases by a factor of 20\% when using the 'S3' class (with a small dependence on spectral index and incidence angle) when the all-sky-background rate increases from $\sim$4.2 Hz ('transient') to $\sim$5.2 Hz ('S3'). 
Our study showed that 'S3' brings improvement over the `transient' class in terms of signal-to-noise ratio ($Signal/\sqrt{Signal+Background}$) above 80 MeV. This improvement depends on the brightness of the burst and is of the order of 15\% for bursts with GRB 080825C characteristics.
As a consequence, the 'S3' event class was used for the spectral analysis of GRB 080825C.
Note that the lower energy threshold does not add any additional systematics compared to the previously used 100 MeV threshold (see section~\ref{sec:specana}).

Finally, the event
selection makes use of the spatial information around the best LAT
localization. The LAT point-spread-function (PSF) has a strong dependence with energy as well as with the conversion point in the tracker. LAT events are thus separated into FRONT (conversion in the upper part of the tracker) and BACK (conversion in the lower part of the tracker) events \citep{atw09} for which separate response functions are provided.
The region of interest (ROI) 
considered in this analysis is energy dependent and based on the $95\%$
containment radius ($PSF95$) and the 95\% LAT error localization ($Err95$):  

$$ ROI(E) = \sqrt{PSF95(E)^2 + Err95^2} $$

To avoid large background contamination, a maximum size is set at 10 and 12 deg for FRONT and BACK events respectively.
In the particular case of GRB 080825C: $ROI(E<200\mbox{ MeV})$ is set to this maximum size and $ROI(\sim 500 \mbox{ MeV}) = 2.9 \mbox{ and } 4.0$ degrees for FRONT and BACK events respectively.

All events resulting from this selection process are
shown in the bottom panel of Figure~\ref{lightcurves}.

\subsubsection{Background estimation}\label{sec:background}

Since the number of events detected by the
LAT is only 15 (of which 13 have an energy $>$80~MeV and can can be used
for spectral analysis),
we need to carefully estimate the expected number of
background events in order to calculate the statistical significance of
our measurements.
The background in the LAT data used for this analysis is
dominated by cosmic rays (CRs), with a small contribution
from extragalactic and Galactic diffuse gamma rays.
Because the earth's limb was $\sim$85~deg from the GRB location, the
contamination from earth-albedo events was negligible and consisted of
events with very poorly reconstructed directions.
The background rate is a function of many parameters and can vary by
more than a factor of two, depending on the observational
conditions. For example, there is a strong dependence of the
CR background rate on the geomagnetic coordinates at the
location of the spacecraft. Furthermore, the
background rate also depends on the burst position in instrument
coordinates, because the LAT's acceptance
varies strongly with the inclination angle.
For these reasons, it is not straightforward to estimate
the expected amount of background during the GRB emission using
off-source regions around the trigger time, since the spacecraft will
have moved to regions of different geomagnetic coordinates,
and the inclination angle of the region of interest will have
changed significantly.  All these effects are properly taken into
account to estimate the background rate for our specific observational
conditions.

Because the two background components (CRs and gamma rays)
have different properties, they are estimated
separately using two different methods. The amount of gamma-ray
background from some direction in the celestial sphere depends only on
the accumulated
exposure in that particular direction. Therefore, this component can be
estimated by simply
scaling the number of gamma rays detected in six months of LAT data,
produced during normal
science operations, by the ratio of the exposure
of the GRB observation over the exposure of the six-month data set.
Similarly to the above, the amount of CR background from some direction
in the celestial sphere
depends on the exposure in that direction. However, unlike the above,
the CR background
also depends on the geomagnetic coordinates at the location of the
spacecraft at each instant of the
observation. Because of the latter dependence, the CR background cannot
be calculated
the same way as the gamma-ray background. Instead,
a Monte Carlo simulation of the GRB observation is performed, in which
parameterizations of the dependence of the CR background rate on the
geomagnetic
coordinates and on the inclination angle are used to
estimate it for each second of the observation. These two methods will
be described in detail below.

The CR-background estimation was based on properties of the LAT data
extracted from a
subset of six months of data selecting when the Galactic plane was far
from the center
of the LAT's FoV ($|B| > 70^{o}$). During such observations, and for
the transient and
S3 data classes, the gamma-ray contribution to the detected signal is
negligible, and the
 detected events can be approximated as being cosmic rays. Specifically,
for the transient
class the gamma-ray contamination in this subset of the data is about
10\%, which consists
of comparable amounts of galactic and extragalactic diffuse emissions
and a negligible contribution
from resolved point sources. For the S3 class, the gamma-ray
contamination is comparable,
although slightly lower. After extracting the dependence of the
CR-background rate on the geomagnetic coordinates at the position of the
spacecraft, we used this dependence to calculate the all-sky
CR-background rate for each second of the GRB observation.
Then, a corresponding
number of simulated events were generated with directions in
instrument-centered coordinates, off-axis
$\theta$ and azimuthal angle $\phi$, selected to match
the observed distribution in the same subset of the six months of LAT
data as used above.
These coordinates were then
converted to equatorial coordinates, using the instrument's pointing
information and were added to a skymap. The simulation procedure
described above
was repeated hundreds of millions of times. The individual skymaps
generated at
each one of the iterations were then averaged to obtain a single skymap
that showed the expected amount of CR background from each direction of the
celestial sphere for the specific GRB observation.

As mentioned above, the gamma-ray component of the background was
estimated by rescaling the number of gamma rays detected during six
months of LAT data
(without applying a cut on the galactic latitude of the LAT pointing
direction this time).
First, a CR background map corresponding to the
six-month period was created by following the procedure mentioned in
the previous paragraph.
Then, an all-particle (both CRs and gamma rays) map was filled with the
directions of
all the events actually detected by the LAT during the six-month period.
Then, the
 estimated CR-background map was subtracted from that all-particle
six-month map to
 produce a residual map that was assumed to contain only gamma rays.
The number of gamma rays
detected during the GRB observation was then calculated by scaling the
residual map with the ratio of the exposures of the two periods (six
months exposure over GRB-observation exposure).

The procedures of estimating the CR and gamma-ray
components of the background described above were repeated for 40
different energy ranges spaced logarithmically from 20~MeV to 300~GeV.
For each energy range, the CR-background estimations used
a different dependence of the CR-background rate on the
geomagnetic coordinates, and $\theta$ and $\phi$ distributions.
Similarly, the gamma-ray background estimations used
different instrument-response functions for the exposure calculations.
The resulting CR and gamma-ray maps
were then added to produce all-particle estimated-background maps (one
for each energy range).
The new maps were then integrated over the energy-dependent regions of
interest to produce the final all-particle background estimates.
The all-sky fraction of gamma-rays in the background appears consistent
with the $|B| > 70^{o}$ gamma-ray fraction, such that it contributes
$\sim10\%$ towards the overall background rate.
The results of this method were tested against actual LAT data for a
variety of durations,
 locations in the celestial sphere, energies, and data classes.
All the distributions of the ratios of the estimated over the
actually-detected
 signals followed a Gaussian distribution, with width about 15\% and
center zero
(no systematic over- or under-estimation of the background).
This accuracy did not depend on the direction of the region of interest
(e.g. its
distance from the Galactic plane) and did not have a strong dependence
($\simeq$$5\%$) on the observation's duration.

\subsubsection{Significance calculation}

Using the `transient' event selection in an energy-dependent analysis region centered around the GBM best localization and no energy selection, we find 15 events
($N_{\rm on}$) between $T_0$ and $T_0 + 35$ s (time interval where we find significant
emission in the NaI detectors).  From the background estimate (Section 2.2.2), the
expected number of counts in the same region and time interval is
$B_{\rm est}=1.3$. 
The gamma-ray contribution to the background estimate for this particular burst was about $8\%$. 
In order to assess carefully the significance of this observation, 
we have used 4 independent statistical methods which are described below.

The first method uses an unbinned likelihood analysis of the LAT data which takes into account the energy-dependent PSF in an event-by-event basis. This method finds a significance of 6.5 comparable to the significances found with more simple counting methods described below. This is due to the fact that our selected events have fairly large PSF and thus spatial information within our ROI is not that constraining. In the case of GRBs where high energy events have been detected, this method is expected to provide the highest significance since it fully takes into account the spatial information for each event.

The second method computes the probability of the null hypothesis being
true (the probability that the observed number of counts in the on-source
region is due to a background fluctuation) in a frequentist approach
that treats the background uncertainty in a semi-Bayesian way
\citep{con03}. Given a certain estimated number of background counts B
during the on-source interval, we compute what is the probability of the
actual on-source measurement $N_{\rm on}$ being consistent with this
value. $P_{sup}(N_{\rm on},B)$ expresses the probability of the on-source
measurement being equal or superior to $N_{\rm on}$ when only statistical
fluctuations are considered:

\begin{equation}
P_{sup}(N_{\rm on},B) = \sum_{N=N_{\rm on}}^{\infty}  \frac{e^{-B} \times B^N}{N!}.
\end{equation}

Because of systematic uncertainties in the background estimation
method, each possible value for the number of background counts is
weighted using a Gaussian distribution with a mean of $B_{\rm est}$ and
a standard deviation of $0.15\times B_{\rm est}$ (since we estimate
our systematics to be about 15\%): ${\rm Gauss}(B)$. We then integrate
over all the possible values of B to compute a weighted probability:

\begin{equation}
P_{\rm null}=
\frac {\int_0^{\infty} dB \mbox{ } {\rm Gauss}(B) \times P_{\rm sup}(N_{\rm on},B)} 
{\int_0^{\infty} dB \mbox{ } {\rm Gauss}(B)}
\end{equation}

We applied this method for the numbers mentioned above: $N_{\rm
on}=15$ events, $B_{\rm est}=1.3$, and we obtained a
null-hypothesis probability of $1.3 \times 10^{-10}$, which corresponds
to a significance of $6.6\;\sigma$.

The third method is fully Bayesian. The question of whether a GRB is
detected by the LAT is analyzed as an on-source/off-source
observation in the time domain.  For the on-source observation the
same parameters are used as in the classical (frequentist) analysis,
$T_{\rm on}= $ 33.0 s of live time (corresponding to a clock time of
35.5 s) and $N_{\rm on}=$15. Because the inclination angle 
of the best source position with respect to the instrument boresight moved slowly in
a 565 s interval around the burst trigger (with $T_{\rm off}=$ 525.0 s
of live time), we could provide an estimate of the off-source
background rate.  During this interval there were $N_{\rm off}=$ 19
counts detected, and the corresponding rate is consistent with the
previous background estimate $B_{\rm est}$.  The spacecraft motion was
favorable for GRB~080825C, allowing an unusually long off-source
interval to determine the background rate.  For some other GRBs,
spacecraft motion may cause the background in the LAT to vary more
quickly, limiting the time for which an off-source interval measures
the same background as that of the on-source interval, and therefore
limiting the applicability of this method.

The Bayesian method assumes that the counts during the background
interval are due to a Poissonian process with rate $b$. To evaluate the
probability that a source has been detected during the on-source
interval, the method compares two hypotheses for the Poisson rate
during the on-source interval: that observed counts are due to the
same background rate $b$, or that they are due to a background plus
source rate $b+s$.  This second hypothesis is insufficiently specified
to quantitatively solve the problem: one must specify some plausible
range for the source rate $s$ using prior data (i.e., not using the
LAT observations). We produce a reasonable estimate for the LAT source counts by extrapolating the time-integrated spectral 
fit of the GBM data to the LAT energy range.  The photon model from the GBM fit is
propagated through the LAT response to predict the number of LAT
counts.  A real LAT observation could have a smaller value than this
estimate because of a statistical fluctuation, or because of a
spectral break between the GBM and LAT energy range. Alternatively, a real
LAT observation could actually exceed this estimate for the maximum
rate if there were an additional and distinct spectral component to the one found 
with the GBM data alone. Nevertheless, this is a
reasonable `prior' estimate for the maximum counts expected in the LAT, 
which for GRB 082525C is $\sim$$60$ counts.  Under the
assumptions described, both \citet{lor92} and \citet{gre05} give
analytic solutions for the probability $P$ that the source is
detected. Using the observational parameters listed above, we find
$1-P =1.2 \times 10^{-8}$ which corresponds to a $5.6\;\sigma$
significance. Moreover, this method can provide the probabilities for the number of events actually originating from the source: all 15 events (30\%), 14 events (35\%), 13 events (22\%), 12 events (9\%), 11 events (3\%), $\leq$10 events (1\%).

Finally we have computed the significance with a fully frequentist
method using the on-source/off-source approach as described by
\citep[equation (17) of][]{lima}. This method yields a significance of 6.4 $\sigma$
for the detecting this burst.

It should be noted that because such search for LAT excess is performed on all GRBs triggered by the GBM and other instruments (when the burst is in the LAT FoV), it is important to consider multi-trials in our analysis. For independent searches as is the case here, the post-trials probability threshold for obtaining a $5 \sigma$ result is $P_{post-trial} = 1- (1- P_{5 \sigma})^{1/N}$ where N is the number of trials and $P_{5 \sigma}$ the $5 \sigma$ probability threshold for a single search ($\sim 5.7 \times 10^{-7}$). 
In the case of GRB 080825C, we searched for LAT excess in $\sim$50 bursts triggered by the GBM which corresponds to a post-trial probability for a $5 \sigma$ results of $P_{5\sigma,post-trial} \sim 1.15 \times 10^{-8}$. This corresponds to a significance of 5.7 which is therefore our threshold for a $5 \sigma$ detection.

The four independent significance computations presented above all yield consistent
results similar or above this threshold. GRB~080825C is therefore the first GRB detected by the LAT
instrument \citep{bou08} at a high significance level.

\subsubsection{Localization}\label{sec:localization}

Due to the strong variation of the LAT PSF with photon energy
\citep{atw09}, the on-ground localization of a source depends strongly
on its spectral shape. For example, 10 photons detected at 100 MeV
will yield an accuracy of $\sim$$1^\circ$, while one single photon
with an energy of 10 GeV will increase the accuracy to $\sim$$0.1^\circ$.

The on-ground localization procedure makes use of the `transient'
class events. It is restricted to the events detected above 100 MeV (which have a good PSF)
in a $15^\circ$ region around the GBM
trigger position (${\rm RA} = 232.2$, ${\rm Dec} = -4.9$). The method
is based on a likelihood ratio test, following the same steps as \citet{Mattox}.
While these authors used a binned likelihood for the analysis of EGRET data, our likelihood function
is unbinned and uses the instrument PSF on an event basis. It also takes into account the various residual backgrounds.
First, the position of the source and its spectrum are left free in the fit, assuming a power-law shape.
Then, in order to compute an accurate localization error, the Test-Statistics (TS, see section \ref{sec:afterglow} for a complete definition)
of the point source spectral fit is computed at each node of a fine map ($5^\circ\times 5^\circ$ for this case, with a bin size of $0.1^\circ$).
Following \citet{Mattox}, the TS values are interpreted in terms of the chi-squared
distribution with two degrees of freedom (the two map coordinates). Figure~\ref{FigLocCL} shows the obtained error contours around the fitted position 
${\rm RA} = 233.9$, ${\rm Dec} = -4.5$, with a 68\%, 90\% and 99\% statistical error radius of 0.8$^\circ$, 1.3$^\circ$ and 2.0$^\circ$, respectively.

At large inclination angles, the systematic error of a source
localization with the LAT is dominated by the slight bias in direction
reconstruction of low-energy photons. This bias is mainly caused by a
trigger effect, which selects those events that scatter downwards and
interact with at least three tracker planes, as required by the
instrument trigger logic \citep{atw09}. The bias is amplified by the
reconstruction efficiency, which is larger for tracks near normal
incidence. Since the PSF is assumed to be centered on average on the true
directions of the gamma rays, this effect translates to a bias in the
fitted position towards smaller inclination angle.

This systematic error has been evaluated in two steps, using both data
and Monte-Carlo simulations. In the first step, we studied the LAT
performance in localizing the Vela pulsar, which is the brightest
source and has a well-determined position from observations at other
wavelengths. We found that the
fitted position obtained from observations where this source was seen
at large inclination angles, is biased towards smaller angles.
This bias is noticeable only when low-energy events (below 1~GeV) are used
and it disappears when high-energy events (above 1~GeV) are included
in the analysis. The agreement found with the prediction of the Monte-Carlo simulation
allows us to evaluate the bias for any burst observation. For instance,
Figure~\ref{FigBiasVsE} shows how the bias varies with energy range for a bright
gamma-ray point source with the same spectral index and inclination
angle as GRB~080916C \citep{abd09}.

In the second step, we used the Monte-Carlo simulation to evaluate the
localization systematic error for GRB~080825C. We produced a simulation of a point
source with the same spectral index ($\sim$$-2.3$) and inclination
angle as GRB~080825C ($\sim$$60^\circ$).  The position obtained with
events between 100~MeV and 570~MeV showed a deviation of
$\sim$$0.6^\circ$ with respect to the input position.  This systematic
error is larger than the one derived for the very bright GRB~080916C
($\lesssim$$0.1^\circ$) because GRB~080825C occurred at a larger
inclination angle and had a maximum photon energy well below 1~GeV.

\section{Data analysis} \label{sec:dataana}

\subsection{Temporal analysis}

\subsubsection{GBM and LAT light curves}

The NaI, BGO and LAT ('S3' selection) light curves are shown in
Figure~\ref{lightcurves}.  No LAT event is seen in coincidence with
the first bright GBM peak. The first 3 LAT events above 80 MeV
are detected in a very short time window, a few seconds after the GBM
trigger, in coincidence with the second GBM peak.  
After a quiet period where no LAT events are detected up to $\sim$$16\;$s, 
4 more events are detected within the GBM $T_{90}$, 
and another 4 events after $T_{90}$ when the
NaI emission has faded close to background level. 
Interestingly the highest energy event,
with an energy of $572 \pm 58$~MeV, is detected at $\sim$$T_0+28\;$s.

To quantify the different features of this GRB (LAT delay, gap, and
extended emission), we have performed Monte-Carlo simulations of the
LAT light curve and estimated the fraction of those simulations that
reproduce these features.  The LAT event distribution was produced
using Poisson statistics for a constant background rate of 0.037$\;$Hz
(see section \ref{sec:background}), an estimated detected signal above
$80\;$MeV of 11.7 events (13 events minus 1.3 estimated background
events) and a temporal probability distribution based on the NaI
light curve from $T_0$ to $T_0+35\;$s. 
The probability of the different features were computed as follow:
\begin{itemize}
\item delay: the fraction of simulated light curves where the first event arrives later than the first actual observed photon (at $T_0+3.252$ s).
\item gap: the fraction of simulated light curves which include a gap in the middle of the light curve with a
width larger than the 12.38 s observed.
\item extended emission: the fraction of simulated light curves where the last 5 events are detected after the timing of the ninth observed event (detected at $T_0 + 26.570$ s).
\end{itemize}

This analysis finds
weak evidence for the possible delay or gap features (with
chance probabilities of 3.4\% and 0.89\%, respectively), but the
evidence for temporally extended emission in the LAT is more significant, at
a 3.7$\;\sigma$ level.

\subsubsection{High-energy afterglow search} \label{sec:afterglow}

We searched for possible afterglow emission up to 13~ks after the trigger time, but did not find any significant
emission. Note that the LAT event detected around $T_0+47$~s is consistent with the
expected background event every $\sim$30 s in the region of interest. 

For the search, we selected the time intervals in which the 10$^\circ$ region of interest centered
on the LAT location (see section \ref{sec:localization}) was in the LAT field-of-view. The burst location exited
the field-of-view 1500 s after the trigger time, re-entered it
$\sim$1 hour later (at $T_0+5200$ s) for $\sim$1600 s, and again $\sim$1.5 hour later (at $T_0+10800$ s) for
$\sim$2400 s.

These periods were split into five time bins
with roughly exponentially increasing durations : 35.5 s to 100 s after the trigger time (just
at the end of the prompt emission), 100 s to 350 s, 350 s to 1500 s,
5200 s to 6800 s and finally 10800 s to 13200 s after the trigger
time. We used the `diffuse' event selection \citep{atw09}, which is adapted to faint source
studies.

In each time bin we computed the Test-Statistics (TS) of a point source with a
power-law spectrum located at the position of the burst. Two spectral
fits were performed using the unbinned maximum likelihood method, one including
only background components (the null hypothesis), the other also including a test point
source (the alternative hypothesis), with two possible spectral indices: $-2.0$ (close to the
spectral index of the prompt emission in the last time bin) and $-1.5$
(i.e., a harder spectrum which could arise from an additional spectral
component at later times).
The Galactic diffuse emission, the isotropic diffuse emission (as
described in \citet{nonGeVxs}) and two likely blazars close to the burst
location (0FGL J1511.2-0536 and 0FGL J1512.7-0905) \citep{BSL} were
included in the background model, and their contribution was estimated
using pre-burst data for the region of interest.
The TS, which is defined as two times the difference of the log-likelihoods between the alternative hypothesis and the null hypothesis,
was derived in each time bin and for each considered spectral index. All TS
values are very close to zero, indicating a null detection.

Since no significant afterglow emission was found, 95\% C.L. upper
limits on the flux of the possible emission were derived,
using a Bayesian method with flat prior \citep{Helene}. This
method was preferred to the likelihood profile \citep{UL} because of
the very low count regime. Since the TS is close to zero, the profile
is not a symmetric parabola and the upper limits derived by the
likelihood profile method do not have proper statistical coverage. The results obtained with the Bayesian method are shown 
in Figure~\ref{fig:ul_afterglow} and in Table~\ref{tab:ul_afterglow}
for both assumed spectral indices. The upper limit found in the first
bin is of the same order of magnitude as the flux derived at the end of the prompt
emission. Then the upper limit decreases when the considered
observation time increases. 

The same method was used to derive 95\%  C.L. upper limits on the flux in
the prompt emission phase where the LAT did not detect any photon
(time bins (a) and (c)). We assumed a spectral index
extrapolating the GBM spectrum. These results are shown in
Figure~\ref{fig:ul_afterglow} and Table~\ref{tab:ul_afterglow} as
well. 

Between bin (e) and bin (h) we assumed that the flux decreases with
time like $t^a$. We used the fitted flux values on each time bin, 
and for bins with TS zero, we set the flux to an
arbitrary low value (this choice has no impact on the following result).
The 68\% C.L. error bars were determined by the
likelihood profile method. Although its coverage is not correct for
the bins of TS zero, this error bar is accurate enough for the present
purpose of fitting the flux decay. It also ensures homogeneity among
the error bars used for the fit. Since the fitted decay slope was not
significant, we used a $\chi^2$ profile method \citep{UL} to derive
upper limits on the flux decay slope for both assumed spectral indices
:  $a<-2.08$ (95\% C.L.) for a spectral index of $-1.5$ and $a<-1.77$
(95\% C.L.) for a spectral index of $-2.0$. 

\subsection{Time-resolved spectroscopy} \label{sec:specana}

\begin{figure}
\includegraphics[width=\textwidth, keepaspectratio]{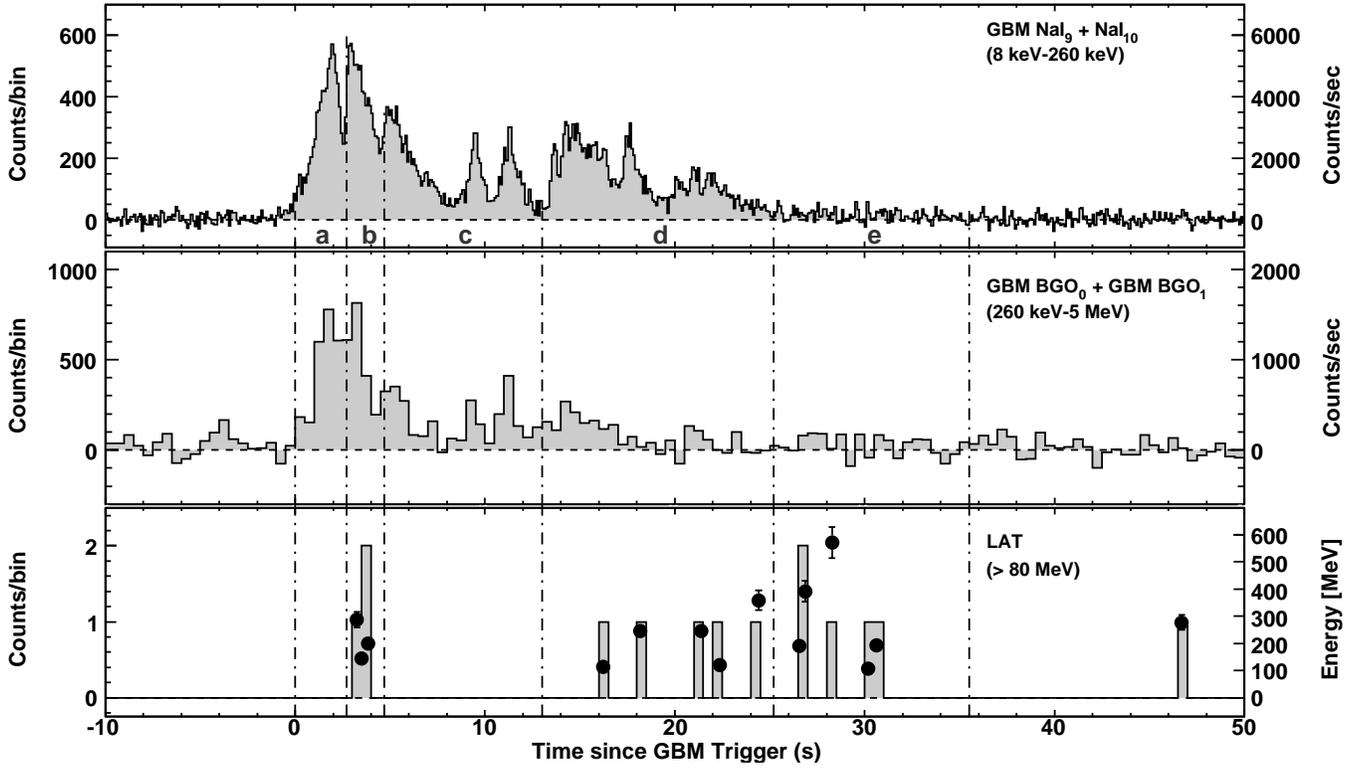}
\caption{Light curves of GRB~080825C observed by the GBM (NaI \& BGO) 
and LAT instruments; top two panels are background subtracted. The LAT
light curve has been generated using events which passed the `S3' event
selection above 80 MeV (which are also the events used for our spectral
analysis). Black dots, along with their error bars (systematic uncertainty in the LAT energy measurement) represent the
$1\;\sigma$ energy range (right y-axis) for each LAT event. 
The vertical dash-dotted lines indicate the time bins used in our time-resolved spectral analysis.}
\label{lightcurves}
\end{figure}

\begin{figure}
\begin{center}
\includegraphics[ width=.8\textwidth, keepaspectratio]{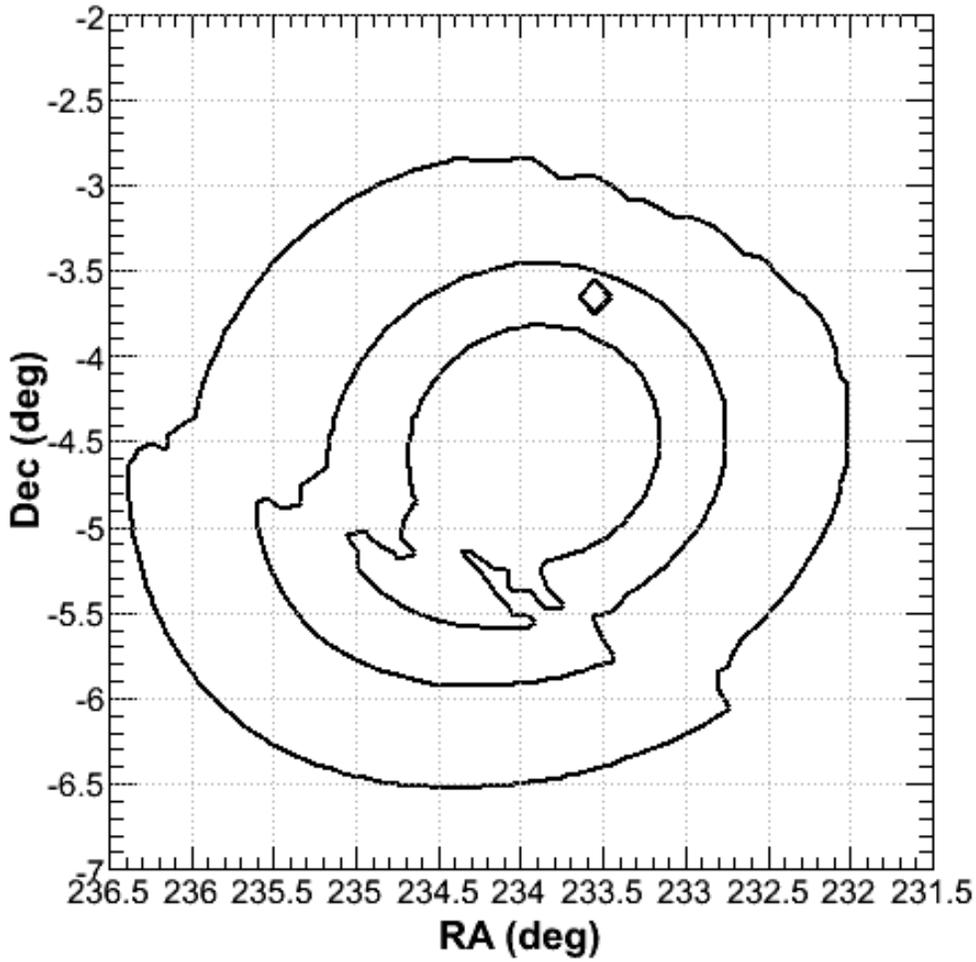}
\end{center}
\caption{LAT on-ground localization for GRB~080825C: RA=233.9, Dec=-4.5. The contours show the containment regions for confidence levels: 68\%, 90\%, and 99\%. Equivalent containment radii can be computed: 0.8$^\circ$ (68\%), 1.3$^\circ$ (90\%) and 2.0$^\circ$ (99\%).}
\label{FigLocCL}
\end{figure}

\begin{figure}
\begin{center}
\includegraphics[width=.8\textwidth, keepaspectratio]{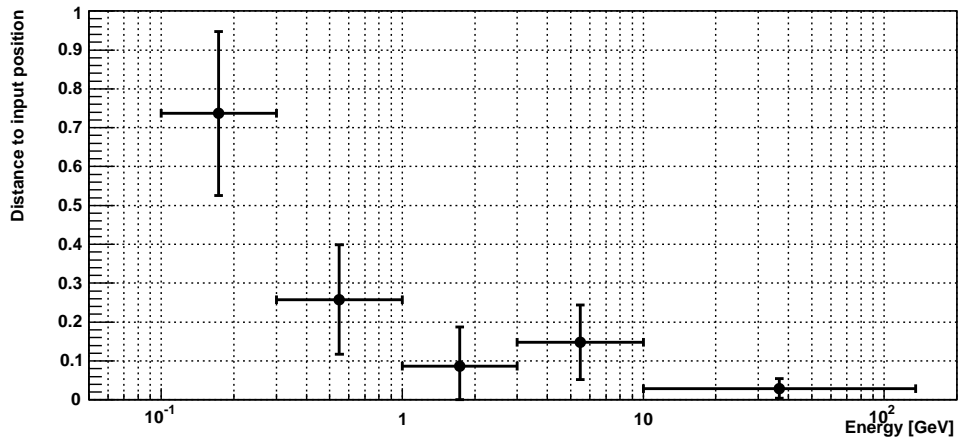}
\end{center}
\caption{Dependence of the localization bias on the energy range of
  the events used. These results are obtained from a simulation of LAT observations of a point
  source similar to GRB~080916C (same slope and inclination
  angle). The use of low-energy events only (below 1 GeV) yields a
  noticeable bias, while no bias is seen when using 
  events only above 1 GeV.}
\label{FigBiasVsE}
\end{figure}

\begin{figure}
\begin{center}
\includegraphics[width=\textwidth, keepaspectratio]{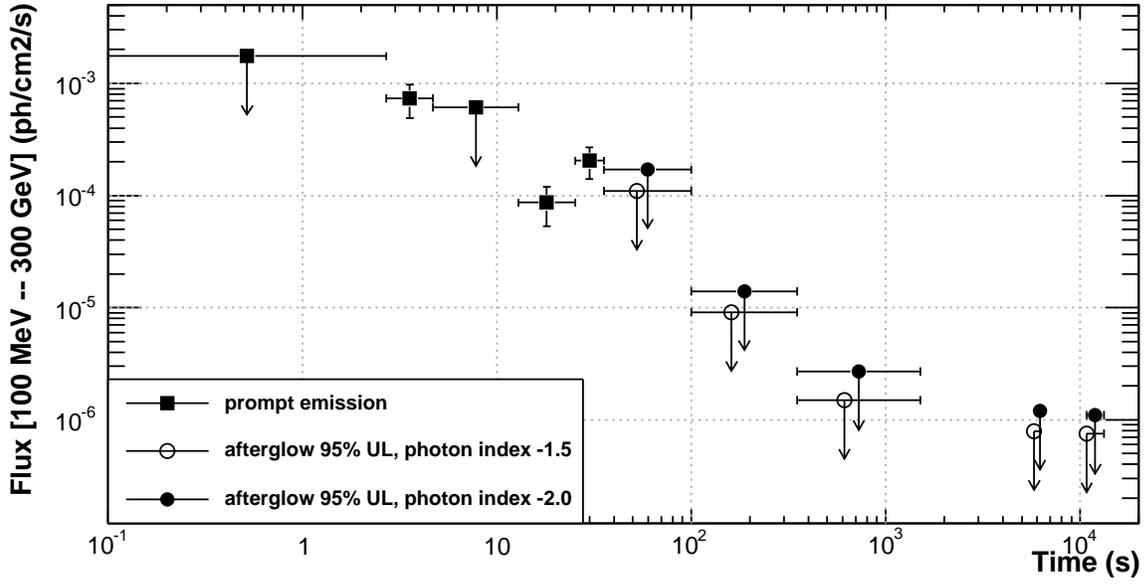}
\end{center}
\caption{Squares: photon fluxes (or 95 \% C.L. upper limits) measured
  during the prompt emission, 
  in the five time bins defined previously. Filled circles (resp. open circles): 95\% 
  C.L. upper limits on the afterglow emission photon flux in each
  considered time bin, assuming a power-law spectral shape of index $-2.0$
  (resp. $-1.5$). \label{fig:ul_afterglow}} 
  \end{figure}

We have performed detailed spectroscopy of the combined GBM and LAT
data, for the whole duration of the burst and also time-resolved
analysis based on the temporal structures observed in the GBM and LAT
light curves.  Figure~\ref{lightcurves} shows the five time intervals
(a to e) adopted for this analysis. The fits were performed with
the spectral analysis software package RMFIT (version 2.5), 
using log-likelihood as the fitting statistic since
chi-squared is less appropriate due to the low number of events at high
gamma-ray energies in this burst.  The variable GBM background is
subtracted for all detectors individually by fitting an
energy-dependent, second order polynomial to background data of $\sim$$300\;$s 
before and $\sim$$300\;$s after the GRB.  For the LAT data, the
`S3' event selection is used (see section \ref{sect:events}).  We used a LAT energy range from 80~MeV to 200~GeV (since
energy bins with no detection still contain useful information) and
adopted a constant but energy-dependent background rate, as described
in section \ref{sec:background}.  For the GBM data we used the standard 128
energy bins of the CSPEC data-type, but only 
the channels above 8~keV in the NaI detectors, above
240~keV in the BGO detectors, and rejecting the overflow channels in
both NaIs and BGOs.

\begin{figure}
\begin{center}
\includegraphics[height=\textwidth, keepaspectratio, angle=90]{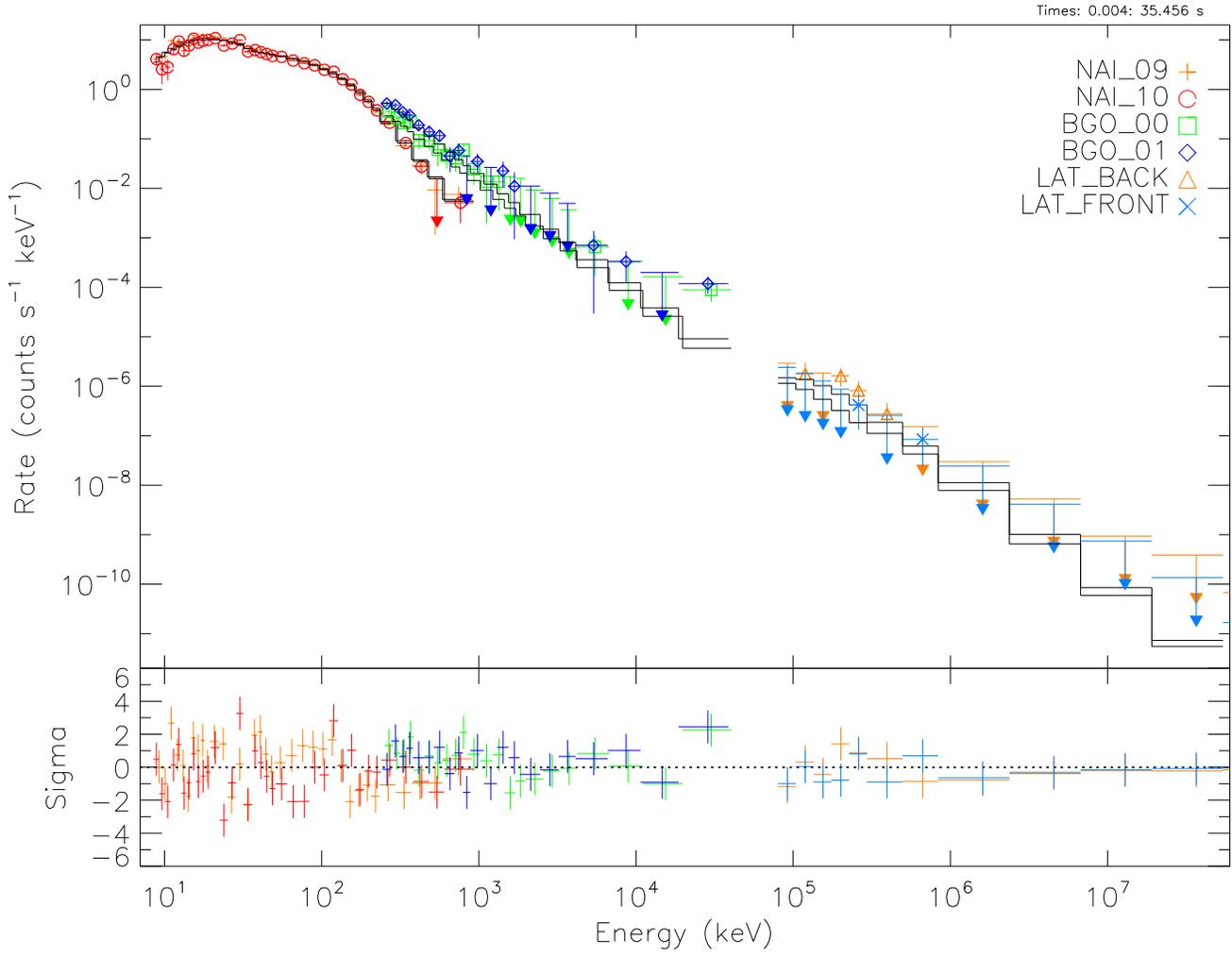}
\end{center}
\caption{Time-integrated ($T_0 + 0.004$ to $T_0 + 35.456$ s) count spectrum of GRB~080825C of the GBM (NaI and BGO) and LAT data. 
The spectrum is well fit by a Band function spanning $\sim$5 decades of energy. 
LAT data has been separated into 'FRONT' and 'BACK' data sets which respectively correspond to events converted in the upper and lower part of the tracker instrument \citep{atw09}.}
\label{intspectrum}
\end{figure}

We considered different empirical models in the spectral analysis: a
simple power law, a power law with a high-energy exponential cutoff,
a Band function \citep{band93} which smoothly connects two power
laws, a Band function with a high-energy exponential cutoff, and a 
Band function with an additional power law. The
high-energy exponential cutoff was implemented by multiplying the
original spectrum by
$\exp(-E/E_{\rm{cutoff}})$. The main results of our combined GBM and
LAT analysis are shown in Tables~\ref{tab:spectres} (spectral parameters) and
\ref{tab:fluxfluence} (flux and fluence in 50 -- 300 keV and 100 -- 600 MeV), 
for which we used the LAT data and responses obtained with the `S3'
cut.  The time-integrated spectrum (shown in Figure~\ref{intspectrum})
and the time bins (b) to (d) are best fit to a significant degree by
the Band function, and thus only the best fit parameters and
associated statistical uncertainties of the Band function are
provided in Table~\ref{tab:spectres} for these time bins.  The best
fit parameters for the Band function are also given for time bin (a), 
although an addition of a possible exponential cutoff is discussed
later in this section.  For time interval (e), however, the spectrum
is adequately described by a single power law, and adding more
parameters does not improve the fit. 
Adding a power law function to the Band function does not improve the time-integrated and time-resolved fits.
Besides spectral fits to the whole GBM \& LAT energy range, we also fitted the GBM data alone 
and propagated the photon model from the GBM fit through the LAT response to obtain the predicted numbers 
of LAT counts. A comparison between the expected and observed numbers of events shows that there is no need 
for an extra emission component besides the Band function (or power law for time bin (e)).

\begin{figure}
\begin{center}
\includegraphics[ width=0.45\textwidth, keepaspectratio]{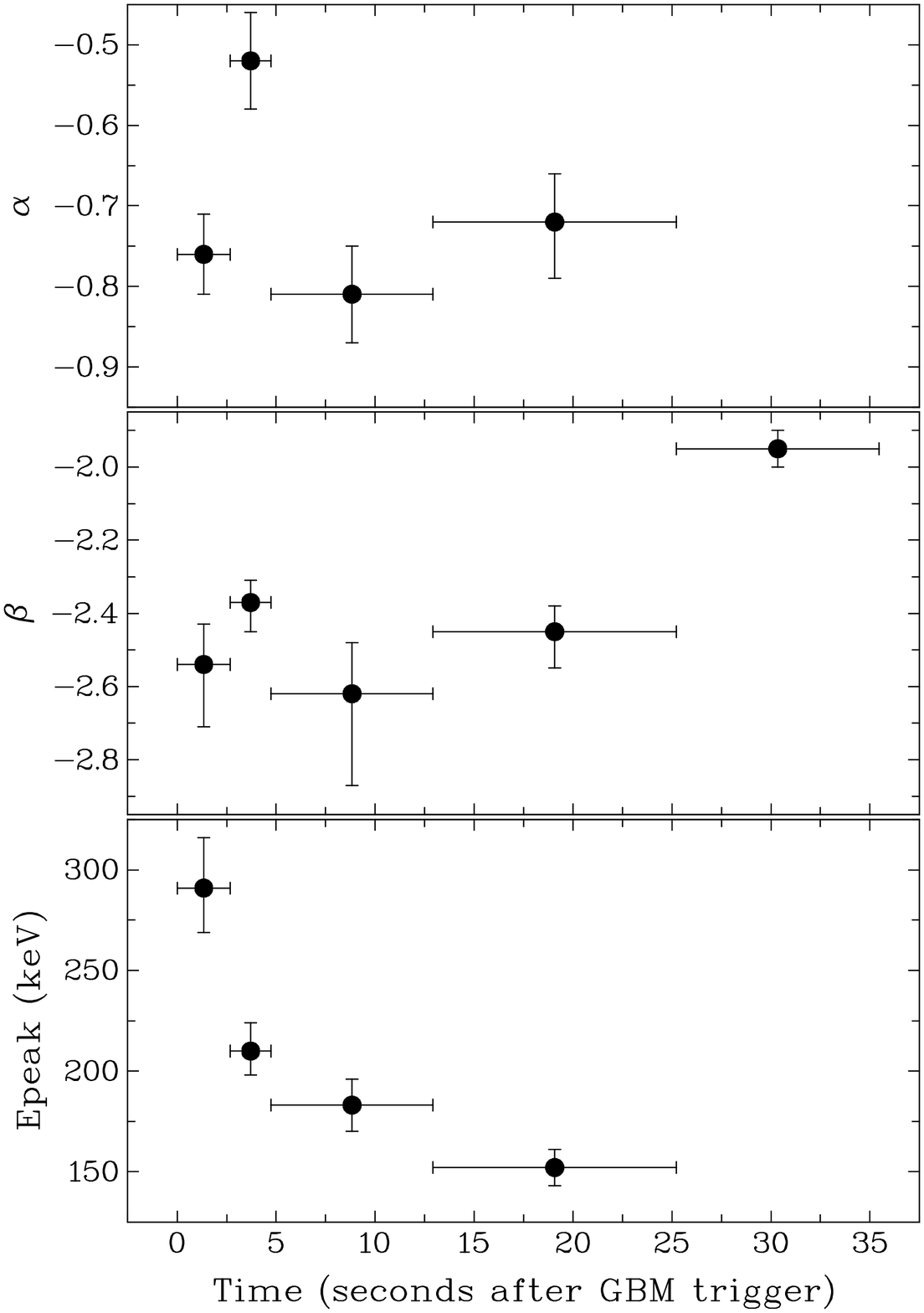}
\hspace{0.1cm}
\includegraphics[ width=0.45\textwidth, keepaspectratio]{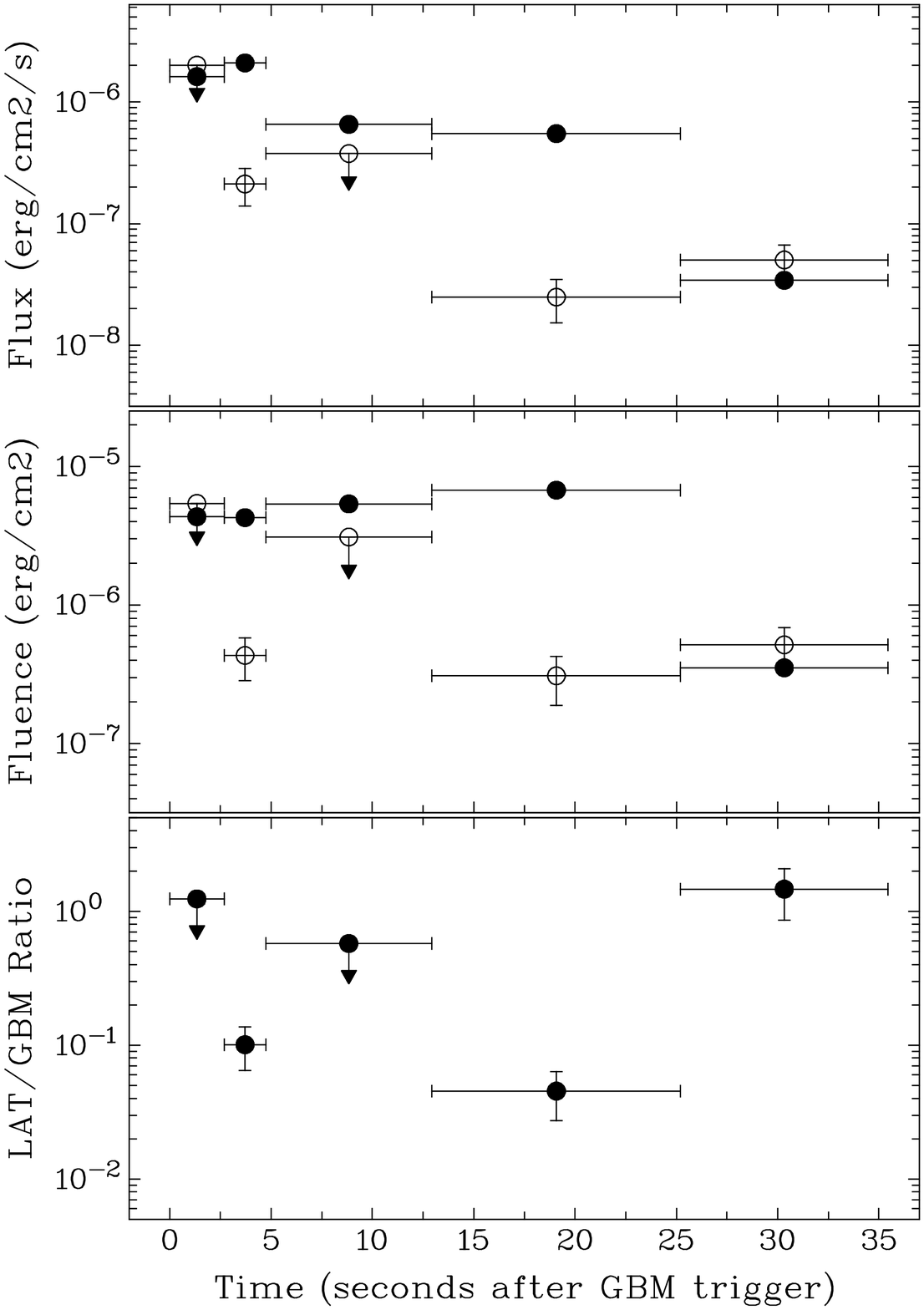}
\end{center}
\caption{{\it Left:} Temporal evolution of the spectral parameters 
for GRB~080825C: the low-energy index $\alpha$ (top), the high-energy
index $\beta$ (middle), and the peak energy $E_{\rm{peak}}$ (bottom). 
The last time bin is adequately fit by a single power-law function, of
which the index is plotted in the middle panel.  {\it Right:} Temporal
evolution of the flux (top) and fluence (middle) in two energy ranges:
50~--~300 keV (solid circles) and 100~--~600 MeV (open circles); and
of the ratio (bottom) between the high-energy and low-energy flux. The
upper limits are given at the $2\;\sigma$ level.}
\label{paramfluxfluence}
\end{figure}

\begin{figure}
\begin{center}
\includegraphics[width=0.98\textwidth, keepaspectratio]{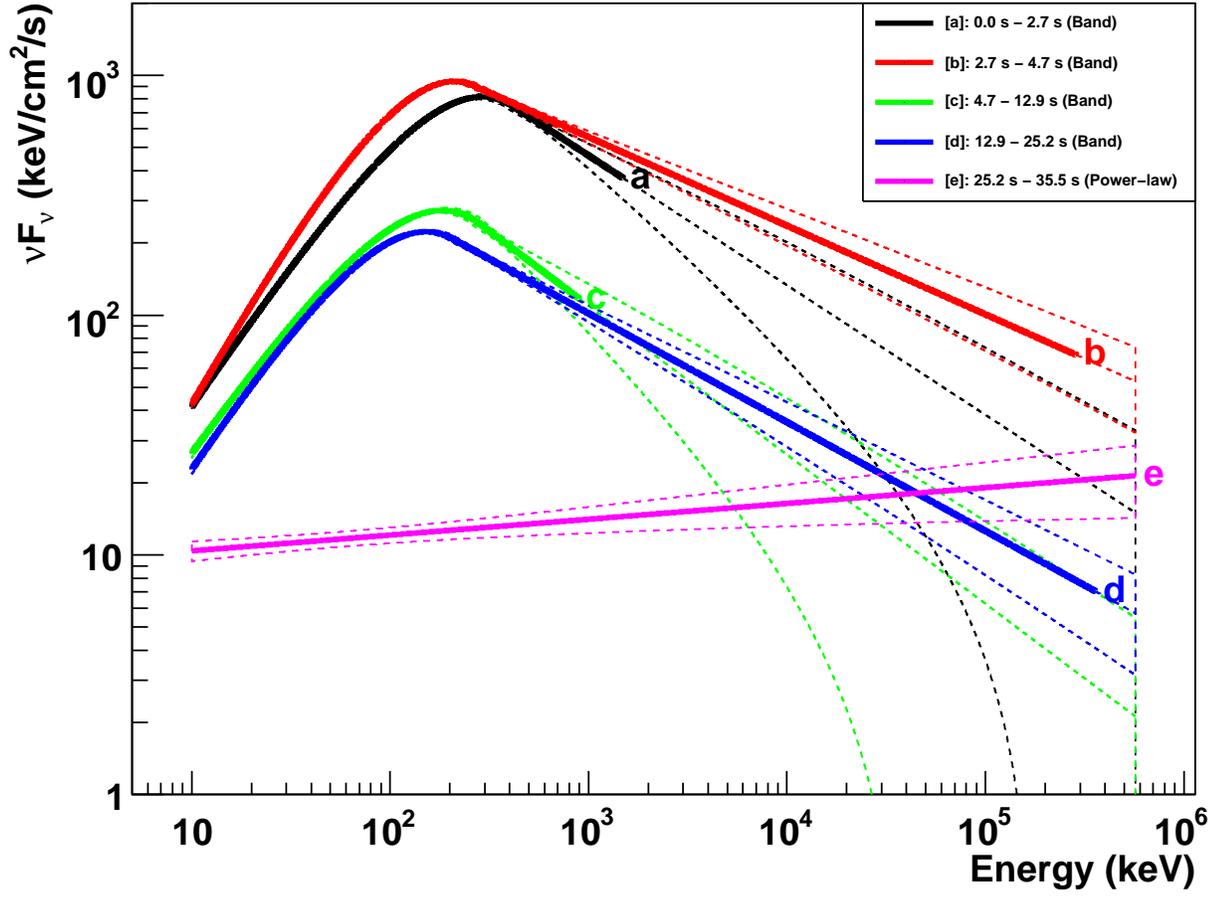}
\end{center}
\caption{The best-fit spectra for time bins (a) to (e) are shown in 
{\it thick solid lines} that reach up to the largest detected photon
energy in each time bin, while the corresponding (same color) {\it
thin dashed} lines represent the $1\,\sigma$ confidence
contours for each fit. Time bins (a) to (d) are fit by a Band function
and time bin (e) is fit by a power-law spectral model.}
\label{fig:multy_spec}
\end{figure}

Time bins (a) to (d) display the
typical hard-to-soft evolution of $E_{\rm{peak}}$ \citep{nor86}, which is the energy at which the Band function peaks in $\nu F_{\nu}$ spectrum, 
starting at almost 300~keV and decreasing to $\sim$$150\;$keV.  Except
for the second interval, the values of the low-energy spectral index,
$\alpha$, and the high-energy spectral index, $\beta$, are constant
within their uncertainties.  The evolution of the spectral parameters,
flux, fluence, and the flux ratio between the two energy ranges, are
shown in Figure~\ref{paramfluxfluence}. The best-fit model spectra for
time bins (a) to (e) are shown in Figure~\ref{fig:multy_spec} along
with their $1\,\sigma$ confidence intervals.

A significant hardening of  the spectrum at high energy is observed  
after $\sim$$25$ sec. From the light curves
(Figure~\ref{lightcurves}) it is clear that while the GBM emission in
time bin (e) is just above the background level, there is significant
emission in the LAT, including the two highest energy events
detected for this GRB.  This spectral hardening is also reflected in
the power-law index of $-1.95\,$$\pm$$\,0.05$ for time bin
(e) (see Table~\ref{tab:spectres}), which is significantly harder than
the values of $\beta$ in the earlier time bins.  Looking at the
evolution of $E_{\rm{peak}}$, one could argue that the spectrum is
affected by curvature at the low-energy end and that a Band function
is not preferred over a single power law due to poor statistics.
However, a spectral fit in interval (e) of all the data above 300~keV
gives a softer but consistent power-law index of
$-2.10\,$$\pm\,$$0.08$.  This $\beta$-value is closer to the
$\beta$-values found for interval (a) to (d), but still significantly
harder.

We have searched for possible departures from a simple Band function in the different time bins by performing a likelihood
ratio test comparing a simple Band function with a Band function
multiplied by an exponential cutoff term, $\exp(-E/E_{\rm{cutoff}})$.  A
significance of $4.3\,\sigma$ was found for an exponential cutoff in
time bin (a) with a cutoff energy around $E_{\rm{cutoff}} = 1.77^{+1.59}_{-0.56}$ MeV (with following Band function parameters: $\alpha \sim -0.57$, $\beta \sim -1.64$, $E_{peak} \sim 211$ keV). We
investigated the dependence of this significance with the systematics
of our instruments and found the strongest effect to be a $\pm\,15\%$
variation in the BGO effective area which can bring the significance
down to $\sim$$3.7\,\sigma$. With 5 time bins, this is not strong
enough to claim the existence of an exponential cutoff.

We have performed a careful investigation of the effect of systematic
uncertainties on the spectral parameters of GRB~080825C.  We
considered the following systematics in the LAT, BGO and NaI
detectors: effective area, energy dispersion and background
subtraction. For the LAT the methodology we use is to modify the
Instrument Response Functions (IRFs) in order to take into account
those various effects.  Systematics on the LAT effective area have
been derived from a study of the Vela pulsar and Earth albedo photons
because it is possible in both cases to extract an extremely pure
gamma-ray sample. Using this sample, the uncertainty on the LAT
effective area has been computed as a function of energy to be 10\%
below 100 MeV, 5\% around 1 GeV and 20\% above 10 GeV. Based on this
estimate, some special IRFs have been created to encompass the extreme
scenarios in overall normalization and slope of the LAT effective
area.  The uncertainty in energy measurement for the LAT is estimated to
be of the order of 5\%.  We adopted a 10\% uncertainty in the NaI and
BGO effective area (both overall normalization and slope).  Finally we
also considered the uncertainty in the choice of off-timing sample for
NaI and BGO background subtraction and found the corresponding
systematics to be negligible.  In each time bin, the error values on
the spectral parameters are found to be similar to or smaller than the
statistical uncertainties reported in Table~\ref{tab:spectres}, except
for the case of time bin (a) and (b) where the systematics on $E_{\rm
peak}$ and the normalization are found to be about two times and three
times larger, respectively. Table~\ref{tab:sys} reports the
systematics found for each parameter of the time-integrated best fit
Band function reported in Table~\ref{tab:spectres} as well as the
predominant systematic effects.  We note that similar trends for the
systematic uncertainties were found for GRB~080916C \citep{abd09},
which had many more LAT events than GRB~080825C.

\section{Discussion} \label{sec:discussion}

\subsection{Theoretical interpretation of GRB 080825C observations}

GRB~080825C is the first GRB detected by the Fermi LAT. Furthermore,
it is the first GRB to show a hint of a time delay between the onset
of the high-energy ($>100\;$MeV) emission relative to the low-energy
(sub-MeV) emission \citep[later seen much more clearly in
GRB~080916C;][]{abd09}. Moreover, there appears to be a local minimum
in the high-energy flux between the initial narrow ($<1\;$s) LAT
spike, at $\sim$$3-4\;$s after the GBM trigger, and the later, much
broader peak (between $\sim$$16\;$s and $\sim$$31\;$s). GRB~080825C is
the first GRB to show this possible feature in the high-energy
emission.

The late broad peak in the high-energy emission has a duration
comparable to its peak time as measured from the GRB trigger time, and
these two timescales are also similar to the duration of the
low-energy emission (which has $T_{90} \sim 27\;$s and $T_{50} \sim 
13\;$s). The fact that these three timescales are comparable is
naturally accounted for if this late broad peak signifies the onset of
the emission from the external (reverse or forward) shocks, if the
reverse shock is (at least mildly) relativistic \citep[see Figure~1
of][]{Sari97}. This occurs around the deceleration time, as the GRB
outflow is decelerated by a reverse shock while it drives a
highly-relativistic forward shock into the external medium. In this
scenario the late broad LAT peak can be either synchrotron
self-Compton (SSC) emission from the reverse shock or external-Compton
(EC) emission from the forward-reverse shock system that forms as the
GRB outflow is decelerated by the external medium \citep[as was
suggested for GRB~941017;][]{GG03,PW04}. In the latter case the seed
synchrotron photons can come from the reverse shock and the
upscattering relativistic electrons from the forward shock, or vice
versa \citep{WDL01}.

In time bin (e), which contains the second half of the broad
high-energy LAT peak while the GBM emission is almost back to
background, the spectrum is well fit by a single power law with a
photon index of $-1.95 \pm 0.05$, which is significantly harder than
the values of the high-energy photon index $\beta$ in all of the
earlier time bins. Moreover, the two highest energy photons in this
GRB are in time bin (e). This may suggest that the late time wide
high-energy peak arises from a separate spectral component to that
responsible for the low-energy emission, which is consistent with an
origin from a distinct physical region (and in particular with an
external shock origin, as mentioned above). If there is still some
contribution from the low-energy spectral component, then the photon
index of the high-energy spectral component, which dominates in time
bin (e) but is sub-dominant earlier on, could be somewhat harder than
the measured value. A rather flat $\nu F_\nu$ spectrum around $\sim$$100\;$MeV 
might suggest that the peak photon energy of this spectral
component is close to this energy range, which may be readily obtained
for SSC from the reverse shock, and possibly for the EC process
described above. Such a spectrum would disfavor SSC from the forward
shock, which typically peaks at TeV energies near the deceleration
time \citep{WDL01,GG03,PW04}.

The rate of the rise in flux up to the wide peak is hard to quantify,
and therefore does not significantly constrain the theoretical
models. The upper limits on the late time high-energy flux imply a
flux decay at least as steep as $\sim t^{-1.7}$. This is
consistent with the steep decay that is expected from the reverse
shock emission after the passage of the reverse shock, but only barely
consistent with forward shock SSC emission \citep[e.g.,][]{SE01} for
fast cooling for a sufficiently soft electron energy distribution,
corresponding to a power-law index of $p \gtrsim 2.6-2.7$. Other
emission mechanisms might also be able to produce a sufficiently steep
flux decay rate.

The possible high-energy cutoff around $E_{\rm{cutoff}} \sim 1.8\;$MeV
in time bin (a), i.e. the first $2.7\;$s from the trigger time, if true, is very interesting, 
as it does not appear to have a good simple explanation.
The lack of a redshift
measurement for GRB~080825C, however, makes it difficult to draw very
strong conclusions, even in this case. In time bin (a), $E_{\rm peak}
\approx 290\;$keV is a factor of $\sim$$6$ lower than
$E_{\rm{cutoff}}$, thus disfavoring a quasi-thermal spectrum. A
possible origin of such a cutoff from intrinsic opacity to pair
production in the source \citep[e.g.,][]{LS01,Bar06,Gra08} might be
hard to reconcile with the non-thermal spectrum and sharp peak in the
GBM light curve in time bin (a), which both imply that the Thomson
optical depth $\tau_T$ of the pairs produced in the source cannot be
$\gg 1$, for any reasonable redshift (that is not $\ll 1$). Such a
reasonable redshift would, in turn, typically imply a bulk Lorentz
factor of the emitting region of $\Gamma \gtrsim 100$, and make it
very difficult to produce $E_{\rm{cutoff}} \sim 1.8\;$MeV,
corresponding to a comoving energy of $E'_{\rm{cutoff}} \sim
18(1+z)(\Gamma/100)^{-1}\;{\rm keV} \ll m_e c^2$. This can be
understood as follows. Producing the observed cutoff through intrinsic
opacity to pair production requires
$\tau_{\gamma\gamma}(E'_{\rm{cutoff}}) \approx 1$. Since the opacity
to pair-production for a photon of energy $E'_{\rm{cutoff}}$ is
produced mainly by photons of energy $\sim E'_{\rm an} \sim (m_e
c^2)^2/E'_{\rm{cutoff}}$ (i.e. near threshold), and the cross-section
to pair production near threshold is of the order of the Thomson
cross-section ($\sigma_T$), then $\tau_{\gamma\gamma}(E') \sim
\tau_T[(m_e c^2)^2/E']$ where $\tau_T(E')$ is the Thomson optical
depth of the pairs that are produced if all the photons of energy near
or above $E'$ pair produce. In particular, taking $E' =
E'_{\rm{cutoff}}$ implies that $\tau_T(E'_{\rm an}) \sim 1$ (i.e. the
Thomson optical depth of the pairs produced by all the photons of
energy $E' \gtrsim E'_{\rm an}$ is of order unity). In our scenario
$\tau_{\gamma\gamma}(E'_{\rm{cutoff}}) \approx 1$ implies that all
photons above $E'_{\rm{cutoff}}$ pair produce, and since they are much
more numerous than the photons above $E'_{\rm an}$ this produces a
very large Thomson optical depth in pairs, $\tau_T(E'_{\rm cutoff})
\sim \tau_{\gamma\gamma}(E'_{\rm an}) \sim (E'_{\rm an}/E'_{\rm
cutoff})^{-1-\beta} \sim (E'_{\rm{cutoff}}/m_e c^2)^{2(1+\beta)} \gg
1$ (where $\beta < -1$ is the high-energy photon index). This is
inconsistent with the observed spectrum and light curve, as mentioned
above, since such a large optical depth would thermalize the spectrum
and suppress the temporal variability. Alternatively, a high-energy
cutoff may reflect the energy spectrum of the accelerated electrons,
but it is not clear why a power-law would extend over a very narrow
range in this case and a much larger dynamical range in most other
cases.

\subsection{Comparison to recent GRBs with high-energy emission}

GRB~080514B was detected by AGILE, triggering on board both the
SuperAGILE X-ray detector (SA) and the AGILE/MCAL detector, and was
observed by the GRID instrument up to 300 MeV \citep{giu08}. The
gamma-ray data above 30 MeV show a significant extended duration with
respect to emission in the hard X-ray and soft gamma-ray energy
bands. A detailed analysis of this component was not possible due to
limited statistics.  Another interesting fact is that in the case of
GRB~080514B, some GRID events are detected when the low-energy
emission has faded beyond detectability both in MCAL and SA
instruments. Similarly, in the case of GRB~080825C, emission in the
NaI detectors has receded to background level when the last LAT events
are detected. In both cases, the behavior of the high-energy emission
in the last part of the prompt emission does not seem to be correlated
with that of the low-energy emission. This suggests that it may
originate from a physically distinct emission region. The similar
behavior observed in the high-energy emission of GRBs 080514B and
080825C may suggest a common explanation.

The longer duration of the high-energy ($>100\;$MeV) emission relative
to the low-energy ($\lesssim 1\;$MeV) emission, and to some extent
(though with much smaller statistical significance) its later onset,
are present not only in GRB~080825C, but also in the two subsequent
LAT GRBs, GRB~080916C \citep{abd09} and GRB~081024B \citep{omo08}.
Therefore, they appear to be common features of at least the first
three LAT GRBs.  Additionally, the onset of the LAT high-energy
emission coincides with the second peak in the GBM low-energy light
curve in all these three LAT GRBs . This has led~\citet{abd09} to
favor an interpretation in which the first and second spikes of the
low-energy light curve originate from two distinct emission regions
with different intrinsic emission spectra, where the second spike has
a harder high-energy photon index $\beta$. This therefore accounts
for the LAT detection of the second spike and non-detection of the
first spike. This interpretation is further supported in the case of
GRB~080916C by the clear change in both the low-energy ($\alpha$) and
high-energy ($\beta$) photon indices between the first and second
peaks of the GBM (low-energy) light curve. This is not clearly seen in
GRB~080825C, possibly since it was not nearly as bright as
GRB~080916C. Other explanations for the delayed onset are possible,
and several other models were also considered by \citet{abd09} for
GRB~080916C. It was hard to clearly distinguish between the different
models on the basis of the available data, despite the extreme
brightness of GRB~080916C (which had an extremely large fluence of
$2.4\times 10^{-4}\;{\rm erg\;cm^{-2}}$ and a record breaking
$E_{\rm\gamma,iso}$ of $8.8\times 10^{54\;}$erg). For GRB~080825C it
is even harder to distinguish between the different possible
explanations.

Temporally extended high-energy ($>100\;$MeV) emission, which lasted
longer than the prompt low-energy (sub-MeV) emission, was detected not
only in GRB~080825C, but also in the other two LAT GRBs mentioned
above (080916C and 081024B), as well as in the AGILE GRB~080514B and
the EGRET GRB~940217~\citep{hur94}. However, GRB~080825C is the only
one so far that shows a hint of a minimum in the high-energy flux
between the early and late high-energy emission, which strengthens the
case for an origin from a distinct physical region, as discussed
above. In GRB~080825C the late time high-energy emission has a harder
photon index than the earlier high-energy emission, which is
consistent with an external shock origin. The opposite is true for GRB
080916C, which favors a somewhat different model such as external
Compton scattering of late time X-ray flare photons by forward shock
electrons. In GRB~080825C the highest-energy photon was detected after
$28\;$s when the low-energy emission was already down to a very low
level, while in GRB~080916C the highest energy photon was detected
after $17\;$s, while the low-energy emission was still very
active. Moreover, the observed duration of the high-energy emission in
GRB~080825C lasted only slightly longer than the low energy emission
(up to $\sim$$31\;$s compared to $T_{90} \sim 27\;$s) while in
GRB~080916C it lasted much longer (more than $\sim$$10^3\;$s, compared
to $T_{90} = 66\;$s). While the high-energy emission in GRB~080916C
was detected for a longer time, partly because it was generally much
brighter than GRB~080825C, it is also possibly because the flux
decayed more slowly at late times (as $t^{-1.2\pm 0.2}$, compared to
steeper than $\sim t^{-1.7}$ in GRB~080825C).

\section{Conclusions}

GRB~080825C is the first GRB detected by the LAT, with 13 events above
80~MeV and a detection significance of $\sim$$6\;\sigma$.  The highest
energy events, up to $\sim$$600\;$MeV, are detected at late times,
$\sim$$25-35\;$s after the GBM trigger, when the emission in GBM has
decreased close to the background level.  The lack of $>1$~GeV
events in the LAT and the large angle of the source to the LAT
boresight result in a localization uncertainty of $1.1^\circ$
(statistical plus systematic) at the $1\;\sigma$ level.

The prompt emission spectrum from both instruments onboard Fermi
covers over 5 decades in energy.  We have performed time-resolved
spectral analysis using the two GBM NaI detectors with the brightest
GRB signal, both GBM BGO detectors, and for the LAT with an event
selection scheme that is optimized for GRB analysis.  We have
carefully taken the energy-dependent backgrounds into account for both
GBM and LAT, and studied the systematic uncertainties in the spectral
analysis. The time-integrated and time-resolved spectra are well fit by the
Band function with a hard-to-soft evolution in the first $25\;$s:
$E_{\rm{peak}}$ evolves from $\sim$$300$ to $\sim$$150\;$keV, the
high-energy power-law index $\beta$ is constant at a value of $\sim$$-2.5$, 
while the low-energy power-law index $\alpha$ is fairly
constant except for the second time bin which contains the first LAT
events.  In the last time bin, $\sim$$25-35\;$s after the GBM trigger
time, the GBM data are barely above background level, and the spectrum
is best fit by a single power law with an index of $\sim$$-2$ which is
significantly harder than the $\beta$ values of the earlier intervals.

The duration and start time of the late broad peak in the high-energy
emission, $\sim$$16-31\;$s after the trigger, suggest that this peak is
emitted by the external reverse or forward shocks, rather than by
internal dissipation within the GRB outflow (e.g., internal shocks or
magnetic reconnection). The relatively fast flux decay after this peak
slightly favors a reverse-forward shock `external' Compton origin
over a forward shock SSC origin. Although the origin and emission
mechanism for this late peak cannot be conclusively determined because
of low number statistics (and the lack of observations at X-ray or
optical wavelengths, due to the poor GRB localization), the external
shock origin is further supported by the change in spectral behavior,
in particular of the spectral index, at these late times.
Observations of more, brighter GRBs with both GBM and LAT will be able
to test this hypothesis.

\acknowledgments
We thank the referee for their detailed and constructive comments.
The $Fermi$ LAT Collaboration acknowledges generous ongoing 
 support from a number of agencies and institutes that have supported 
 both the development and the operation of the LAT as well as scientific data analysis.  
 These include the National Aeronautics and Space Administration and the 
 Department of Energy in the United States, the Commissariat \`a l'Energie Atomique 
 and the Centre National de la Recherche Scientifique / Institut National de Physique Nucl\'eaire 
 et de Physique des Particules in France, the Agenzia Spaziale Italiana 
 and the Istituto Nazionale di Fisica Nucleare in Italy, the Ministry of Education, 
 Culture, Sports, Science and Technology (MEXT), High Energy Accelerator 
 Research Organization (KEK) and Japan Aerospace Exploration Agency (JAXA) in Japan, 
 and the K.~A.~Wallenberg Foundation, the Swedish Research Council 
 and the Swedish National Space Board in Sweden.

Additional support for science analysis during the operations phase 
is gratefully acknowledged from the Istituto Nazionale di Astrofisica in Italy.

J.G. gratefully acknowledges a Royal Society Wolfson Research Merit Award. 
A.J.v.d.H. was supported by an appointment to the NASA Postdoctoral Program at the MSFC, 
administered by Oak Ridge Associated Universities through a contract with NASA.

\clearpage

\begin{table}
\begin{center}
\caption{95\% C.L. upper limit on the photon flux in the LAT energy range
  derived by a Bayesian method with a flat prior. In time 
  bins (a) and (c) of the prompt emission the continuation of the
  spectrum fitted on GBM data was assumed. In time bins (f) to (j) two
  different cases were studied: a spectral index of $-2.0$ (close to the
  spectral index of the prompt emission in the last time bin) and of $-1.5$
  (i.e. a harder spectrum which could arise from an additional spectral 
  component at later times). \label{tab:ul_afterglow}}
\begin{tabular}{|c|c|c|}
\hline
Time & Spectral & Photon flux above 100 MeV \\
Bins (s) & index & 95\% C.L. upper limit (10$^{-6}$ph cm$^{-2}$ s$^{-1}$) \\
\hline
\hline
a : 0.00 -- 2.69 & $-2.54$ & 1750 \\
c : 4.74 -- 12.93 & $-2.62$ & 610 \\
\hline
f : 35.5 -- 100. & $-2.0$ / $-1.5$ & 170 / 110 \\
g : 100. -- 350. & $-2.0$ / $-1.5$ & 14 / 9.1 \\
h : 350. -- 1500. & $-2.0$ / $-1.5$ & 2.7 / 1.5 \\
i : 5200. -- 6800. & $-2.0$ / $-1.5$ & 1.2 / 0.79 \\
j : 10800. -- 13200. & $-2.0$ / $-1.5$ & 1.1 / 0.75 \\
\hline 
\end{tabular}
\end{center}
\end{table}

\begin{table}
\begin{center}
\caption{Time-integrated and time-resolved spectral analysis results for GRB\,080825C. 
Band function best fit parameters are provided for all spectra, except for
time interval (e) which is adequately fit by a single power law.}
\label{tab:spectres}
\renewcommand{\arraystretch}{1.2}
\begin{tabular}{|c|c|c|c|c|} 
\hline
Time Range (s) & A (10$^{-3}$ $\gamma$ cm$^{-2}$ s$^{-1}$ keV$^{-1}$) & $\alpha$ & $\beta$ & $E_{\rm{peak}}$ (keV) \\
\hline\hline
a:   0.00 -- 2.69 &   $75^{\,+6}_{\,-5}$      & $-0.76 \pm 0.05$ & $-2.54^{\,+0.11}_{\,-0.17}$ & $291^{\,+25}_{\,-22}$ \\
b:   2.69 -- 4.74 & $138^{\,+13}_{\,-11}$ & $-0.52 \pm 0.06$ & $-2.37^{\,+0.06}_{\,-0.08}$ & $210^{\,+14}_{\,-12}$ \\
c:   4.74 -- 12.93 &   $44 \pm 4$      & $-0.81 \pm 0.06$ & $-2.62^{\,+0.14}_{\,-0.25}$ & $183 \pm 13$ \\
d: 12.93 -- 25.22 &  $47^{\,+5}_{\,-4}$      & $-0.72^{\,+0.07}_{\,-0.06}$ & $-2.45^{\,+0.07}_{\,-0.10}$ & $152 \pm 9$ \\
e: 25.22 -- 35.46 &    $1.2 \pm 0.1 \mbox{ (at 100 keV)} $ & N.A. & $-1.95 \pm 0.05$ & N.A. \\
\hline
      0.00 -- 35.46 &   $37 \pm 2$     & $-0.79 \pm 0.03$ & $-2.42^{\,+0.04}_{\,-0.05}$ & $198 \pm 8$      \\
\hline
\end{tabular}
\end{center}
\end{table}

\begin{table}
\begin{center}
\caption{Flux and fluence in the 50 -- 300 keV and 100 -- 600 MeV energy ranges 
for the time intervals and spectral parameters presented in Table
\ref{tab:spectres}.  The upper limits are given at the $2\;\sigma$
level.}
\label{tab:fluxfluence}
\renewcommand{\arraystretch}{1.2}
\begin{tabular}{|c|c|c|c|c|} 
\hline
Time & Flux$_{\,50-300\,\rm{keV}}$ & Flux$_{\,100-600\,\rm{MeV}}$ & Fluence$_{\,50-300\,\rm{keV}}$ & Fluence$_{\,100-600\,\rm{MeV}}$ \\
bin & (10$^{-7}$ erg s$^{-1}$ cm$^{-2}$) & (10$^{-8}$ erg s$^{-1}$ cm$^{-2}$) & (10$^{-6}$ erg cm$^{-2}$) & (10$^{-7}$ erg cm$^{-2}$) \\
\hline\hline
a     & $16.2\pm 0.2$   & $<20.1$  & $4.35\pm 0.05$ & $<5.39$ \\
b     & $20.9\pm 0.3$   & $21.1\pm 7.2$    & $4.28\pm 0.06$ & $4.32\pm 1.47$ \\
c     & $6.57\pm 0.11$ & $<3.78$ & $5.38\pm 0.09$ & $<3.09$ \\
d     & $5.50\pm 0.07$ & $2.50\pm 0.97$ & $6.76\pm 0.09$ & $3.07\pm 1.19$ \\
e     & $0.34\pm 0.03$ & $5.03\pm 1.68$ & $0.35\pm 0.03$ & $5.15\pm 1.72$ \\
\hline
Full & $5.98\pm 0.04$  & $4.35\pm 0.88$ & $21.2\pm 0.1$  & $15.4\pm 3.1$ \\
\hline
\end{tabular}
\end{center}
\end{table}

\begin{table}
\begin{center}
\caption{Systematic uncertainties as well as the predominant effect for each spectral parameter of the time-integrated best fit Band function reported in Table \ref{tab:spectres}.}
\label{tab:sys}
\renewcommand{\arraystretch}{1.2}
\begin{tabular}{|c|c|c|c|c|} 
\hline
 & Norm & $\alpha$ & $\beta$ & $E_{\rm{peak}}$ (keV) \\
\hline
systematic error & $\pm 15 \%$ & $\pm 0.03$ & $\pm 0.03$ & $\pm 8$ \\
\hline
predominant systematics & NaI eff. area & NaI eff. area & BGO eff. area & NaI \& BGO eff. area \\
\hline
\end{tabular}
\end{center}
\end{table}

\end{document}